\documentclass[apsrev4,pra,superscriptaddress,notitlepage]{revtex4-1} 
\usepackage{graphicx}
\usepackage{amssymb,calc}
\usepackage{amsmath,amsfonts,amssymb}
\usepackage{graphicx}
\usepackage[colorlinks=true, allcolors=blue]{hyperref}
\usepackage[version=4]{mhchem}

\begin{document}
\title{Multilayer volume holographic gratings from Bayfol\textregistered HX: light and neutron optical characteristics}
\author{Saba Shams Lahijani}
\affiliation{Faculty of Physics, University of Vienna, Boltzmanngasse 5, 1090 Wien, Austria}
\author{Tobias Jenke}
\affiliation{Institut Laue-Langevin, Grenoble, France}
\author{Christian Pruner}
\affiliation{Faculty of Natural and Life Sciences, Department of Chemistry and Physics of Materials, University of Salzburg, Jakob-Haringer-Strasse 2a, A-5020 Salzburg, Austria}
\author{J\"urgen Klepp}
\affiliation{Faculty of Physics, University of Vienna, Boltzmanngasse 5, 1090 Wien, Austria}
\author{Martin Fally}
\affiliation{Faculty of Physics, University of Vienna, Boltzmanngasse 5, 1090 Wien, Austria}
\email{martin.fally@univie.ac.at}
\date{\jobname}
\hypersetup{pdfauthor={Shams Lajijani,Jenke, Pruner, Klepp, Fally},pdftitle={Multilayer volume holographic gratings from BayfolHX: light and neutron optical characteristics}}

\newcommand{\bcr}{[b_c\Delta\rho]_1}
\newcommand{\bay}{Bayfol\textregistered HX~}

\begin{abstract}
During the last decade a number of volume holographic media have been investigated that could serve not only as diffractive optical elements (DOEs) for light but also for slow neutrons. In this contribution we discuss the light optical properties of a stack of two gratings separated by an optically inert slice recorded in a \bay photopolymer. While the refractive-index modulation of the gratings for light is remarkable, the corresponding neutron optical analogue is, so far, in the medium range of other materials investigated. We therefore aim at possible improvements which are discussed in this manuscript.
\end{abstract}

\maketitle
\section{INTRODUCTION}
\label{sec:intro}  
The application of volume holographic media in optical devices is ubiquitous. Suitable materials should exhibit a number of properties such as high sensitivity, the wavelength response, no/low shrinkage, no/low temperature dependence, easy processing, and last not least: low cost. In general the central figure of merit is the light-induced first order saturated refractive-index modulation $\Delta n_{1,L}$. While for light optical DOEs this value is maximized and might reach up to $\Delta n_{1,L}\approx 0.034$ \cite{Bruder-Polymers17} the corresponding neutron optical parameter $\bcr$ - usually referred to as coherent scattering length density modulation - originally was explored only for a few standard photopolymers \cite{Rupp-prl90,Matull-epl91,Rupp-p97,Havermeyer-phb00}. Note, that for neutrons the refractive index at thermal and cold wavelengths is close to $1$ so that refraction or reflection effects are marginal and DOEs are the only choice when it comes to optical devices. Due to the tiny $\bcr\approx 0.1 \mu\textrm{m}^{-2}$ with a resulting $\Delta n_{1,N}=\lambda_N^2/(2\pi)\leq 2\times 10^{-8}$ at a typical cold neutron wavelength of a nanometer gratings have to be very thick (a few millimeters) to result in a detectable diffraction efficiency $\eta_1\approx 0.01$. By employing deuterated polymers and assembling three of these improved diffractive elements in a triple-Laue configuration the first cold neutron interferometers based on such artificial holographic gratings could be designed \cite{Schellhorn-phb97,Pruner-nima06}. One disadvantage when using such thick gratings is their high angular selectivity. This results in rocking curves $\eta_1(\theta)$, i.e., the angular dependence of the diffraction efficiency, that are extremely narrow, oscillations of the diffraction efficiency being averaged out due to beam divergence, and the alignment of three of those gratings being excessively demanding. Therefore, a race for photosensitive materials with a high $\bcr$ was started during the last decade. First attempts with holographically produced thinner gratings (a few tens of micrometers) with holographic polymer dispersed liquid crystals paved the way\cite{Fally-prl06,Drevensek-Olenik-spie07,Fally-joa09}. Next, various nanoparticle-polymer composites of excellent quality proved to be extremely efficient, reliable and versatile \cite{Fally-oex21,Tomita-prappl20,Tomita-jmo16,Klepp-apl12.02,Klepp-apl12.01,Fally-prl10} and are the top-materials for neutrons at present with $\bcr\leq 10~\mu\textrm{m}^{-2}$, whereas photopolymer-ionic liquid composites have so far not reached the same level in $\bcr$ \cite{Flauger-polymers19}. 

In this contribution we focus on an off-the-shelf material which is a commercially available photopolymer of highest optical quality (\bay) \cite{Bruder-Polymers17}. The latter has been characterized extensively and applied for light optical purposes \cite{Chakraborty-photmdpi22,Bruder-SPIE21,Sevilla-photmdpi2021,Blanche-materialsmdpi20,Borisov-stjit20,Bruder-SPIE20a,Chrysler-SPIE18, Francardi-SPIE18,Marin-Saez-oex16,Bruder-SPIE15b,Bruder-SPIE15a,Berneth-SPIE11,Yu-ao17}. Applications of DOEs for other fields such as cold atom trapping or in astronomy were envisaged, too \cite{Marin-Saez-AppliedEnergy19,Tempone-Wiltshire-oex17,Zanutta-spie17,Zanutta-omex16}.


\section{PREPARATION OF THE VOLUME HOLOGRAPHIC GRATINGS}
\label{sec:prep}
For preparing the DOEs we used a \bay film. It consists of a stack of three layers: (1) a transparent substrate layer \textsf{S} of thickness $d_S\approx 50~\mu\textrm{m}$, (2) the ultimate photosensitive layer \textsf{P} with a nominal thickness $d_P\approx 16~\mu\textrm{m}$, and on top (3) a cover layer \textsf{C}. Thus the light-sensitive layer is sandwiched between the substrate layer, which is optically inactive and intended to act as a support, while the cover layer is simply for protection purposes and to be removed before recording. Only recently, films were developed to also allow for a removal of the substrate \cite{Bruder-SPIE20}. However, as we will show below one can benefit from the substrate layer by simply using it as a well defined spacer.

After having removed the cover we prepared two different types of samples from two sheets on top of each other. Thus the stack consists in total of four layers being arranged in two different variants:
\begin{enumerate}
 \item a configuration \textsf{SPPS} which actually is a single polymer layer of thickness $2d_\textsf{P}$. Here, both substrate layers are irrelevant. \label{en:spps}
 \item a configuration \textsf{SPSP} (triple layer). Here, one optically inactive substrate layer \textsf{S} acts as a spacer between two photosensitive layers \textsf{P} and the second substrate layer \textsf{S} is irrelevant. \label{en:spsp}
\end{enumerate}

The grating(s) were recorded using a standard two-wave mixing setup. Two coherent s-polarized beams of wavelength $\lambda_w=514~\textrm{nm}$ (Coherent Genesis CX-514 SLM) and a total dosage of about 60 mJ are brought to interference under an angle of about $2\Theta=62.8^\circ$. The samples with their surface normal vector $\hat \sigma$ parallel to the bisectrix of the beams are placed in the interference region. The resulting spatially varying cosinusoidal light intensity pattern records an unslanted, holographic transmission grating in the photosensitive layers with a spacing of $\Lambda=\lambda_w/(2\sin{\Theta})\approx 493~\textrm{nm}$. In other words: the refractive index is periodically modulated along a direction $x$ perpendicular to $\hat\sigma$:
\begin{equation}
n(x)=n_0+\sum_{s\geq 
\\0} \Delta n_s\cos{\left(sGx+\varphi_s\right)} 
\end{equation}
Here, $G=2\pi/\Lambda$ is the spatial frequency, $\Lambda$ the grating spacing, $n_0$ is the average refractive index of the film, $\Delta n_s$ and $\varphi_s$ are the real valued Fourier-coefficients and the relative phases of order $s$, respectively.

It was shown that the \bay material also exhibits strong higher harmonics $\Delta n_s$ for $s>1$ due to the non-local photo-polymerization driven diffusion \cite{Bruder-spie15}, which can successfully explain and describe the experimental findings by using a sophisticated model \cite{Sheridan-josaa00,ONeill-ao02,Kelly-josab05,Gleeson-josab09.02,Gleeson-josab09.01,Gleeson-josab10,Bruder-SPIE10}.

\section{SINGLE LAYER VOLUME HOLOGRAPHIC GRATING}
\label{sec:SGGS}
Here we present the experimental results for configuration \ref{en:spps} as defined in Sec.~\ref{sec:prep}. We characterize the grating by measuring the angular dependence of the diffraction efficiency $\eta_{\pm 1}(\Theta)$ for light and neutrons, respectively. As the layers are arranged \textsf{SPPS} with a single grating \textsf{G} recorded in layers \textsf{PP} we expect that the properties should be the same as for a single layer of twice the corresponding thickness ($2d_\mathsf{P}=d_\mathsf{G}\approx 32~\mu\textrm{m}$).
\subsection{Light optical properties}
\label{sec:SGGS_L}
\begin{figure} [ht]
   \begin{center}
   \begin{tabular}{c} 
   \includegraphics[width=\columnwidth]{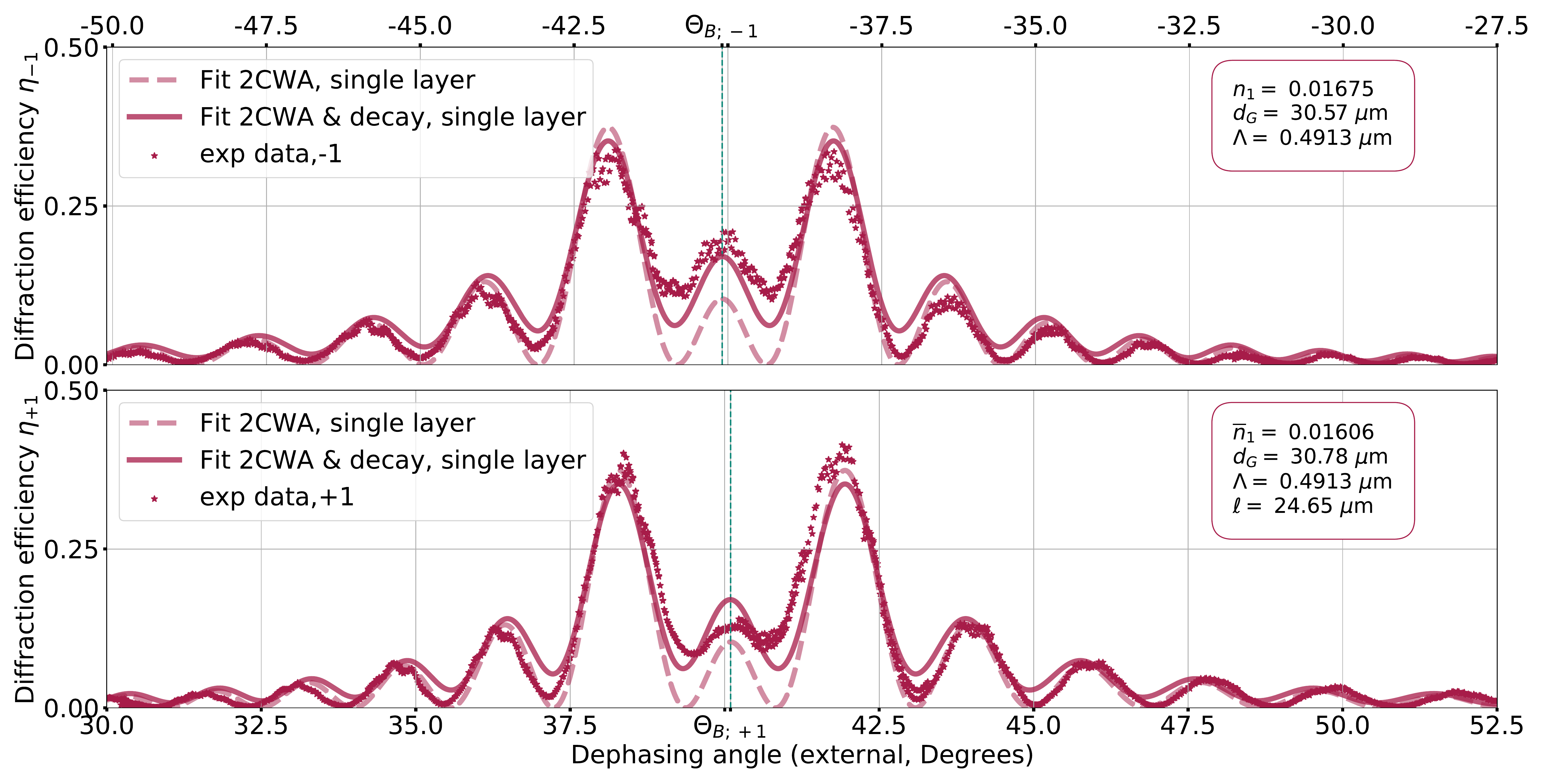}
   \end{tabular}
   \end{center}
   \caption[light1Lred] 
   { \label{fig:1L-red} 
Angular dependence of the diffraction efficiency $\eta_{\pm 1}(\Theta)$ for a configuration \textsf{SPPS}, i.e., a single transmission grating with doubled thickness. Readout wavelength is $\lambda_r=632.8~\textrm{nm}$. Dashed lines are a fit to the data using a first-order coupled wave analysis (parameters see inset, top right) and solid lines show the fit for Uchida's model of an exponentially decaying refractive-index modulation $n_1(z)$ with $\ell$ being the decay length (parameters see inset, bottom right).}
   \end{figure}

The angular dependence of the diffraction efficiency was measured using an s-polarized laser beam of wavelength $\lambda_r=632.8~\textrm{nm}$. The expected external Bragg-angles are $\Theta_{\pm 1}=\pm 39.9^\circ$. Fig. \ref{fig:1L-red} shows $\eta_{\pm 1}(\Theta)$ together with a fit using a first-order coupled wave analysis employing the beta-value method for the boundary conditions \cite{Uchida-josa73,Prijatelj-pra13}.

As expected the Bayfol grating shows excellent properties and the fit provides fairly reliable values for the parameters. An obvious discrepancy of the model with the experimental data with respect to the minima next to the Bragg angle $\Theta_B$ can be explained either by a decay of the refractive-index modulation $\Delta n_1(z)$ along the depth of the sample \cite{Uchida-josa73} and/or  holographic scattering \cite{Tomita-oex20}. Fits to the data according to Uchida's model \cite{Uchida-josa73} are shown in solid lines.
\subsection{Neutron optical properties}
\label{sec:SGGS_N}
To probe the tentative usability of gratings recorded in \bay for neutron optical purpose we performed neutron diffraction experiments at the very cold neutron beamline PF2/VCN at the Institut Laue-Langevin, Grenoble, France. A small-angle neutron diffraction setup was established. By using a set of appropriate slits at a distance, the divergence of the beam was adjusted to about 1 mrad. The VCN-beam has a broad wavelength distribution ranging from 3 to 10 nm, however, the particular distribution strongly depends on the chosen geometry\cite{Oda-nima17}. Here, due to lack of additional time-of-flight data,  we assume that it can be well approximated with an exponentially modified Gaussian\cite{Blaickner-nima19} having a mean wavelength of $\overline{\lambda}=5.5~\textrm{nm}$. Similarly, as in the light optical analogue, the angular dependence of the diffraction signals was measured. Diffracted neutrons are detected downstream at a 2D-detector with a pixel size of $2\times 2~\textrm{mm}^2$. The beams travel from the entrance slit to the detector in a He-tube to minimize unwanted neutron-air scattering. We accounted for any (constant) background counts $I_B$ in the fitting procedure of the diffraction efficiency by assuming $\eta_{\pm 1}=(I_{\pm 1}-I_B)/(\sum_jI_j-jI_B)$.

\begin{figure} [ht]
   \begin{center}
   \begin{tabular}{c} 
   \includegraphics[width=2\columnwidth/3]{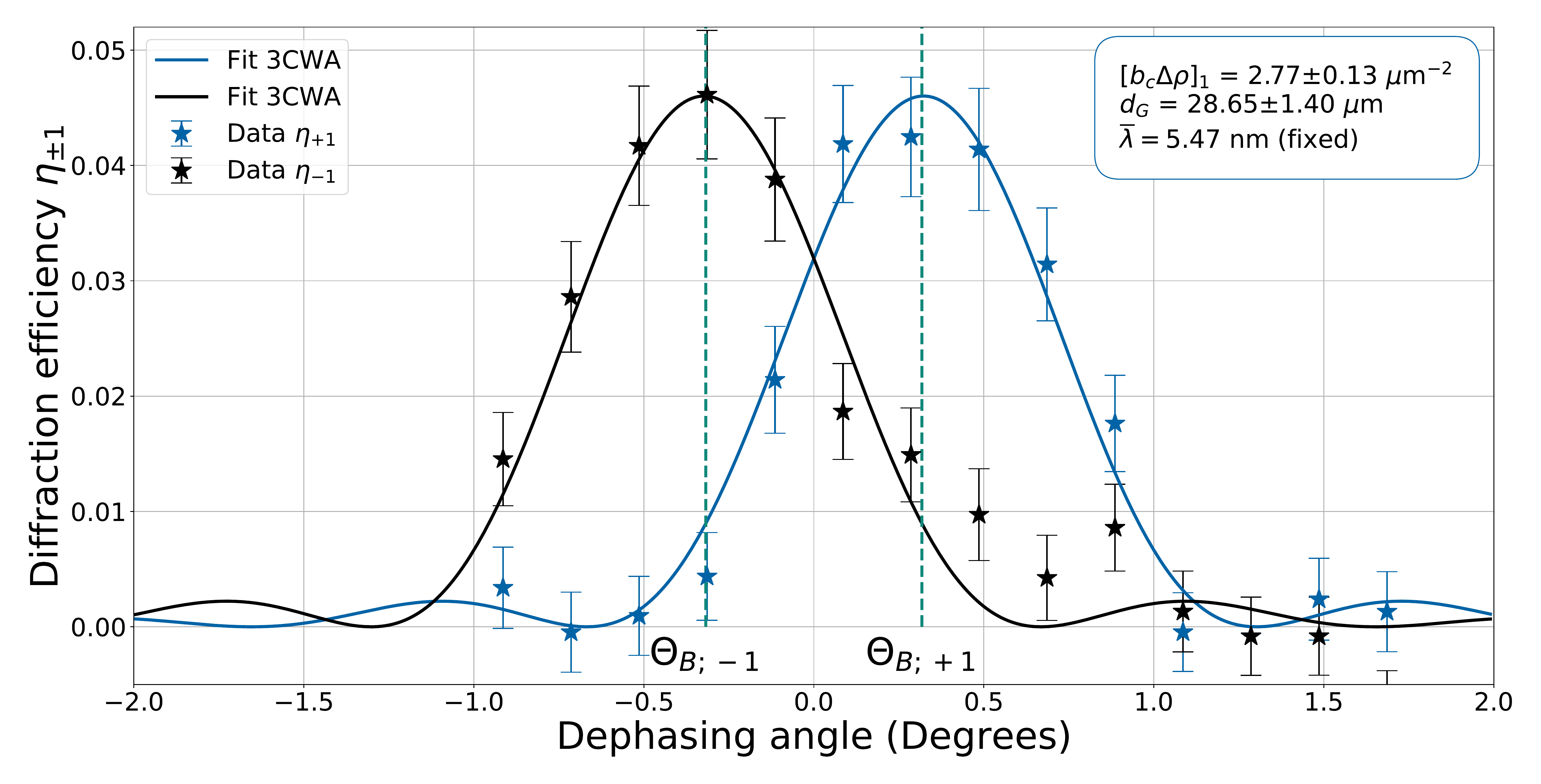}
   \end{tabular}
   \end{center}
   \caption[1Lneut] 
   { \label{fig:1L-neut} 
Angular dependence of the diffraction efficiency $\eta_{\pm 1}(\Theta)$ for a configuration \textsf{SPPS} measured with very cold neutrons of mean wavelength $\overline{\lambda}=5.47~\textrm{nm}$. Solid lines are a fit to the data using a first-order coupled wave analysis with three waves (3CW), i.e., $\pm$ first and zero order diffraction.}
   \end{figure}

The experimental neutron rocking curves are shown in Fig.~\ref{fig:1L-neut}. Due to the small ratio $\overline{\lambda}/\Lambda$ first order Bragg-angles are as small as $\Theta_{\pm 1}=\pm 0.32^\circ$. The half-width of the peaks, however, depends on the ratio $\Lambda/d\approx 0.9^\circ$ which in turn means that peaks of $\pm 1$ orders overlap over a broad angular range. Therefore, at least a three-wave coupling analysis must be applied to make a proper fit to the data \cite{Klepp-jpcser16}. The results of fitting this model are drawn as solid lines in Fig.~\ref{fig:1L-neut}.

\section{MULTILAYER VOLUME HOLOGRAPHIC GRATING}
\label{sec:SGSG}
In this section we present the experimental results characterizing the diffraction properties for the configuration \ref{en:spsp}, \textsf{SPSP}, for light optics and simulations for the neutron optics case derived from the parameters obtained in Sec.~\ref{sec:SGGS_N}.
\subsection{Light optical properties}\label{sec:SGSG_L}
\begin{figure} [ht]
   \begin{center}
   \begin{tabular}{cc} 
   \includegraphics[width=\columnwidth]{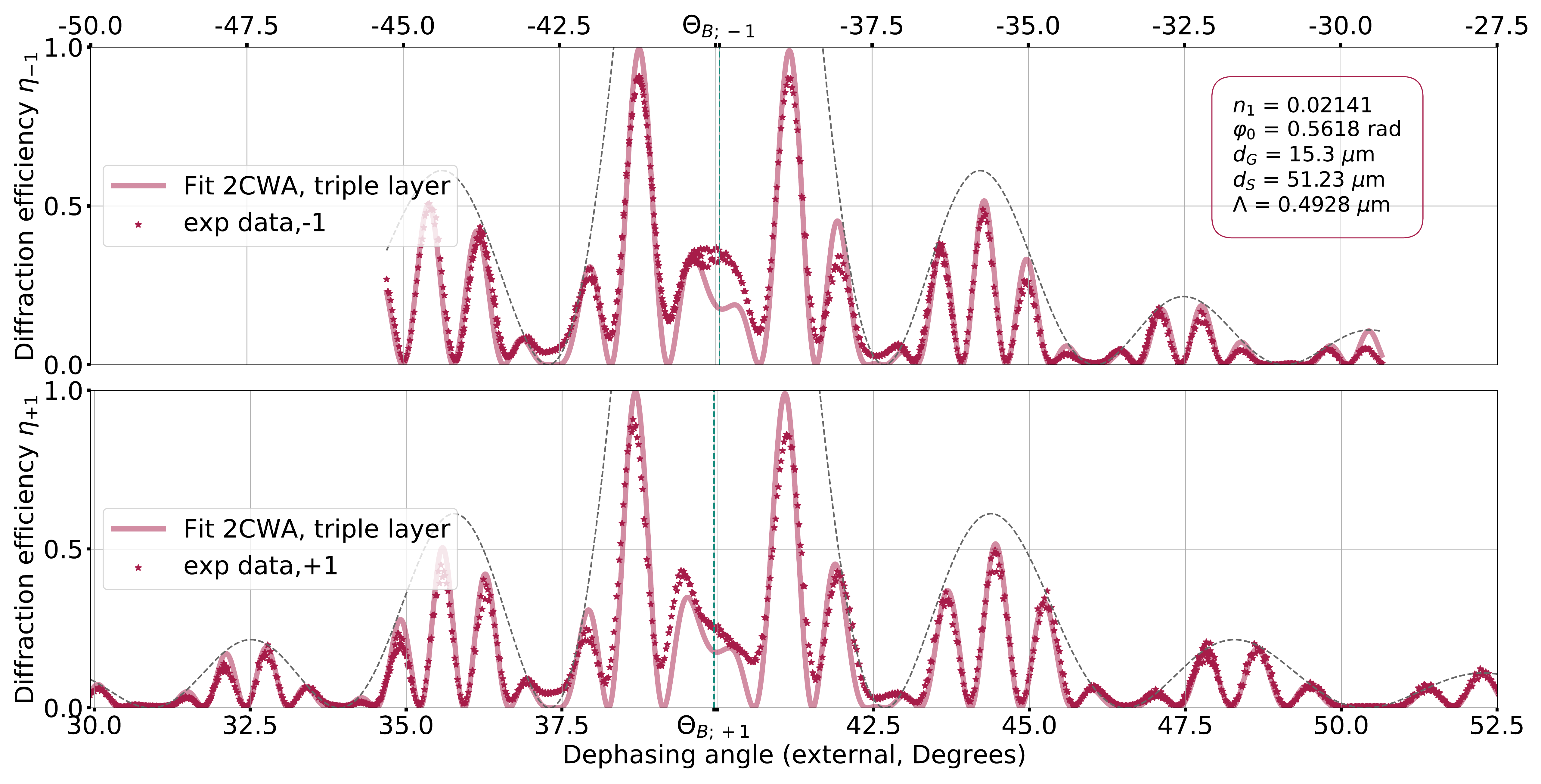}
   \end{tabular}
   \end{center}
   \caption[light1Lred] 
   { \label{fig:3L-red1} 
Angular dependence of the diffraction efficiency $\eta_{\pm 1}(\Theta)$ for a configuration \textsf{SPSP}, i.e., a layer of two gratings separated by a spacer of thickness $d_S$. Readout wavelength is $\lambda_r=632.8~\textrm{nm}$. Solid lines are a fit to the data using a first-order coupled wave analysis \cite{Au-jmo87}. We assume identical gratings in both photopolymer layers (see Eq.~\ref{eq:2CW-3L1}).}
   \end{figure}

The angular dependence of the light optical diffraction from an effective triple layer \textsf{SPSP} configuration, where each \textsf{P} is transformed to a grating \textsf{G}, are shown in Fig. \ref{fig:3L-red1}. We observe slow and rapid oscillations of the angular dependence of the ($\pm$first order) diffraction efficiencies. The slow oscillations marked by the dashed envelope function are due to the single grating layer of a thickness $d_G\approx 16~\mu\textrm{m}$. The rapid oscillations, in contrast, depend on the thickness $d_S\approx 50~\mu\textrm{m}$ of the sandwich-substrate layer. Such multilayers are well known as reflective elements in light optics, e.g., for spectral filters or ultra-high reflection Bragg mirrors\cite{Cole-NaturalPhotonics13}. Actually two stacks of them constitute a Laue-Laue geometry microscale interferometer, a device that is interesting also in neutron optics.

Modelling the angular response of diffraction efficiency for a system of layers \textsf{SPSP} can be done by employing the amplitude transmittance approach for each layer as given in Ref.~\citenum{Au-jmo87}. The total diffraction amplitudes are found by multiplying the input amplitudes by a product of characteristic matrices for each layer. For the present case we first assume that the two photopolymer layers, which were exposed in a single recording process, have identical refractive-index modulations $n_1$ and thicknesses $d_\mathsf{G}$. However, we allow for a phase-shift $\varphi_0$ between the gratings. Then for the triple layer system $\eta_{\pm 1}$ is given by
\newcommand{\sinc}{\textrm{sinc\,}}
\begin{eqnarray}
 \label{eq:2CW-3L1}
 \eta_{\pm 1}&=&\left(2\nu \sinc{\Phi}
 \left[
 \cos(\xi_\mathsf{S}+\varphi_0/2)\cos\Phi-\xi_\mathsf{G}\sin(\xi_S+\varphi_0/2)\sinc{\Phi}
 \right]\right)^2\\
 \nu&=&\frac{\beta n_{\pm 1}d_\mathsf{G}}{2n_0\sqrt{c_0c_{\pm 1}}}\\
 \xi_{\ell}&=&\frac{d_{\ell}\beta(c_0-c_{\pm 1})}{2}; \ell=\mathsf{G,S}\\
 \Phi&=&\sqrt{\nu^2+\xi_\mathsf{G}^2}\\
 \nonumber
 c_0&=&\cos(\theta),\quad c_{\mp 1}=\sqrt{1-(\sin\theta\pm G/\beta)^2}
\end{eqnarray}
The dephasing angles $\theta$ are those measured in the medium. The experimental data show a lifting of the minima near the Bragg peaks (indicated by a dashed line) while this cannot be described by the model used here (see Fig.~\ref{fig:3L-red1}). The dashed lines show the envelope function. 



\begin{figure} [ht]
   \begin{center}
   \begin{tabular}{cc} 
   \includegraphics[width=2\columnwidth/3-2mm]{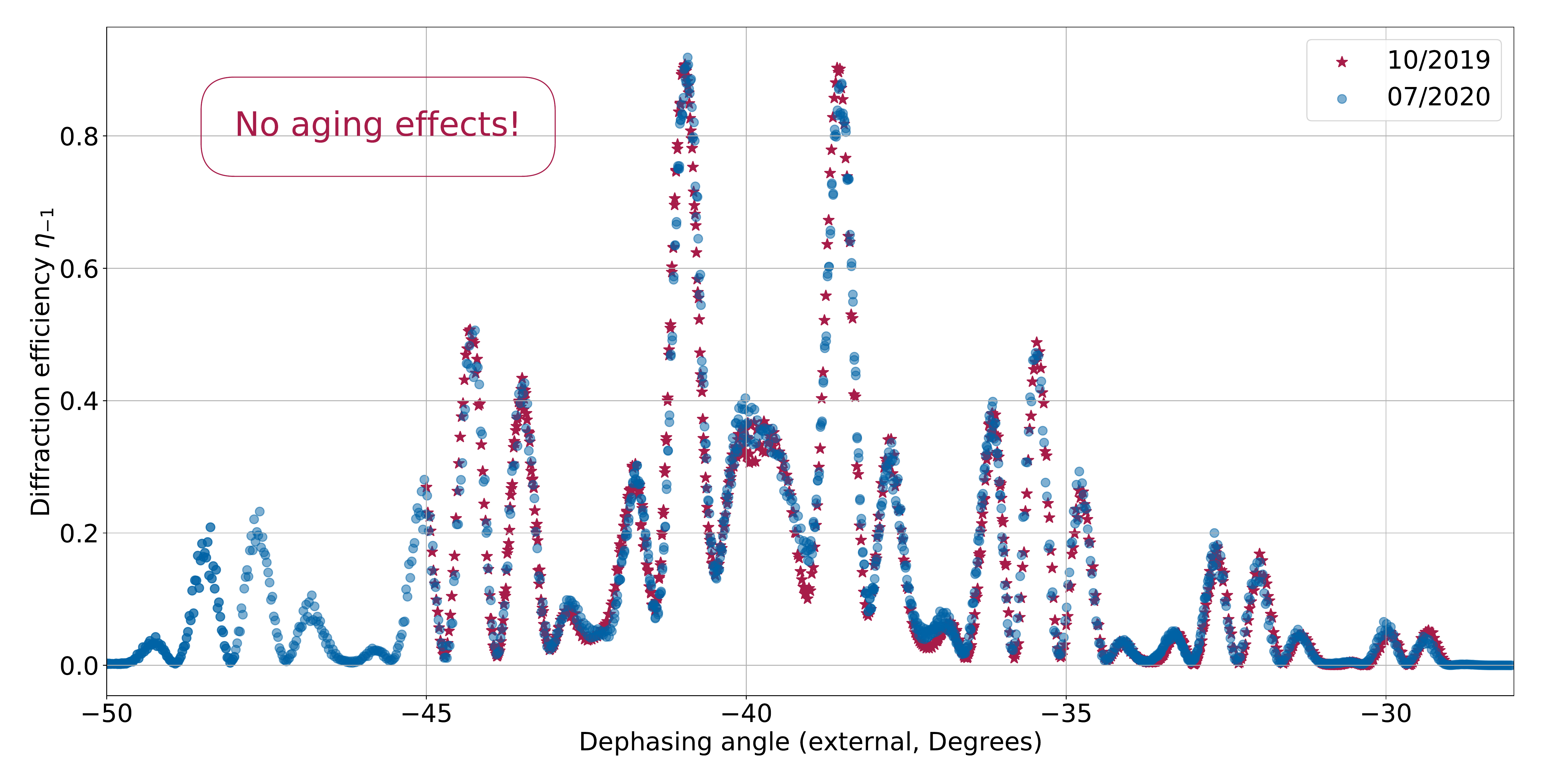}
   \end{tabular}
   \end{center}
   \caption[light1Lred] 
   { \label{fig:ageing} 
Angular dependence of the diffraction efficiency $\eta_{-1}(\Theta)$ for a configuration \textsf{SPSP} directly after recording the stack (10/2019) and nine months later (09/2020). Readout wavelength is $\lambda_r=632.8~\textrm{nm}$.}
   \end{figure}

Another important figure of merit for gratings from Bayfol is their temporal stability. In a previous neutron interferometer with an extremely demanding setup the gratings became misagligned within a few months \cite{Pruner-nima06}. Therefore, we checked the recorded Bayfol gratings for changes over time and found that even after nine months ageing did not occur (see Fig.~\ref{fig:ageing}).

\subsection{Neutron optical properties - simulations}
\label{sec:SGSG_N}
In this subsection we will simulate the outcome of neutron diffraction properties for the cases of a stack of two films resulting effectively in either a single layer \textsf{SPPS} or a triple layer \textsf{SPSP}, and for the cases of a stack of three films resulting in triple layers, too. For the latter case, additional configurations are:
 \begin{enumerate}
  \setcounter{enumi}{2}
  \item \textsf{PSSPPS} or \textsf{SPPSSP}\label{en:psspps}
  \item \textsf{SPPSPS} or \textsf{SPSPPS}\label{en:sppsps}
 \end{enumerate}

For simulating the neutron optical properties for each of these configurations we make use of data and fitting results shown in the previous sections. We assume that the wavelength distribution is the same as the one used in Fig.~\ref{fig:1L-neut}. In addition we recognize that the refractive-index modulation for light and the triple layers is by a factor $r=2.141/1.675\approx 1.28$ larger as for the single layer hologram. Thus we scale up the measured $\bcr$ for neutrons and the single layer to estimate that for the triple layer. This results in a $\bcr\approx 3.54~\mu\textrm{m}^{-2}$ for the simulations. The average wavelength is set to $\overline{\lambda}=5.47$ nm, using an exponentially modified Gaussian distribution \cite{Blaickner-nima19}.
The thicknesses and the phase are chosen as obtained from the fits (see Fig.~\ref{fig:1L-neut} and \ref{fig:3L-red1})

\begin{figure}[ht]
   \begin{center}
   \begin{tabular}{cc} 
   \includegraphics[width=2\columnwidth/3-2mm]{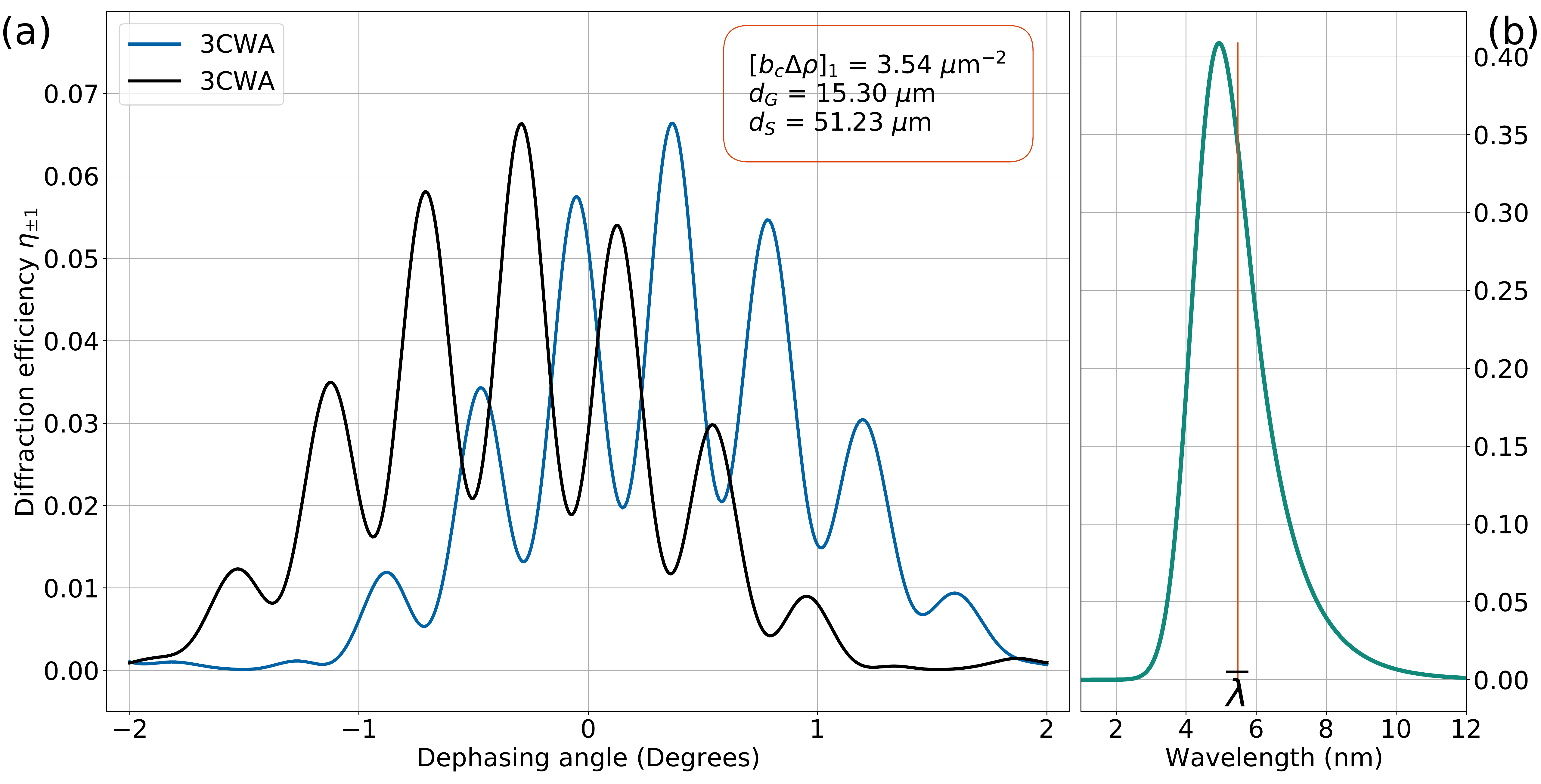}
   \end{tabular}
   \end{center}
   \caption[light1Lred] 
   { \label{fig:sim_spsp} 
 (a) Simulated angular dependence of the diffraction efficiency $\eta_{\pm 1}(\Theta)$ for neutrons for a configuration \textsf{SPSP} using a feasible exponentially modified Gaussian wavelength distribution as shown in (b). The average wavelength is $\overline{\lambda}=5.47~\textrm{nm}$.}
   \end{figure}

Fig.~\ref{fig:sim_spsp} shows the simulated angular dependence of the diffraction efficiency for very cold neutrons (compare Table~\ref{tab:parameters}) from a triple layer (\textsf{SPSP}, see Fig.~\ref{fig:3L-red1} for light diffraction data). A summary of the decisive holographic grating parameters $n_1$ and the thicknesses $d_G$ for light or neutrons in both configurations can be found in Table~\ref{tab:parameters}.

\begin{table}[th]
\caption{Summary of grating characteristic parameters as obtained from fits for light and neutrons, respectively, for configurations as defined in Sec.~\ref{sec:prep}. Grating spacing $\Lambda\approx 493~$nm. For neutrons the refractive-index modulation $n_1$ is calculated from the fitted $\bcr$ according to the relation $n_1=\bcr\lambda^2/(2\pi)$. }
\label{tab:parameters}
\begin{center}       
\begin{tabular}{|l|l|c|}
\hline
\multicolumn{1}{|c|}{Configuration} & \multicolumn{1}{|c|}{Parameters}  & \multicolumn{1}{|c|}{Model} \\\hline
\multicolumn{3}{|c|}{Light, $\lambda_r=632.8$~nm} \\ \hline
Single layer \textsf{SPPS}, Sec.~\ref{sec:SGGS_L} & $n_1=(1.675\pm 0.004)\times 10^{-2}$ & 2CWA\cite{Yeh-93} \\
                                        &$d_G=(30.57\pm 0.04)~\mu\mathrm{m}$ & \\
                                        &$\overline{n}_1=(1.606\pm 0.007)\times 10^{-2}$ & 2CWA decay\cite{Uchida-josa73}\\
                                        &$d_G=(30.78\pm 0.09)~\mu\mathrm{m}$ &\\\hline
Triple layer \textsf{SPSP}, Sec.~\ref{sec:SGSG_L} & $n_1=(2.141\pm0.005)\times 10^{-2}$ & 2CWA Multilayer\cite{Au-jmo87}\\
                                            & $d_G=(15.30\pm 0.01)~\mu\mathrm{m}$ & Eq. \ref{eq:2CW-3L1}\\
                                            & $d_S=(51.23\pm 0.02)~\mu\mathrm{m}$ & \\\hline\hline
\multicolumn{3}{|c|}{Neutrons, $\overline{\lambda}=5.47$ nm} \\ \hline
Single layer \textsf{SPPS}, Sec.~\ref{sec:SGGS_N} &$n_1=(1.32\pm 0.06 )\times 10^{-5}$ & 3CWA\cite{Klepp-jpcser16} \\
                                         &$d_G=(28.7\pm 1.4)~\mu\mathrm{m}$ & \\\hline

\end{tabular}
\end{center}
\end{table}

\begin{figure}[ht]
   \begin{center}
   \begin{tabular}{cc} 
   \includegraphics[width=3\columnwidth/4-2mm]{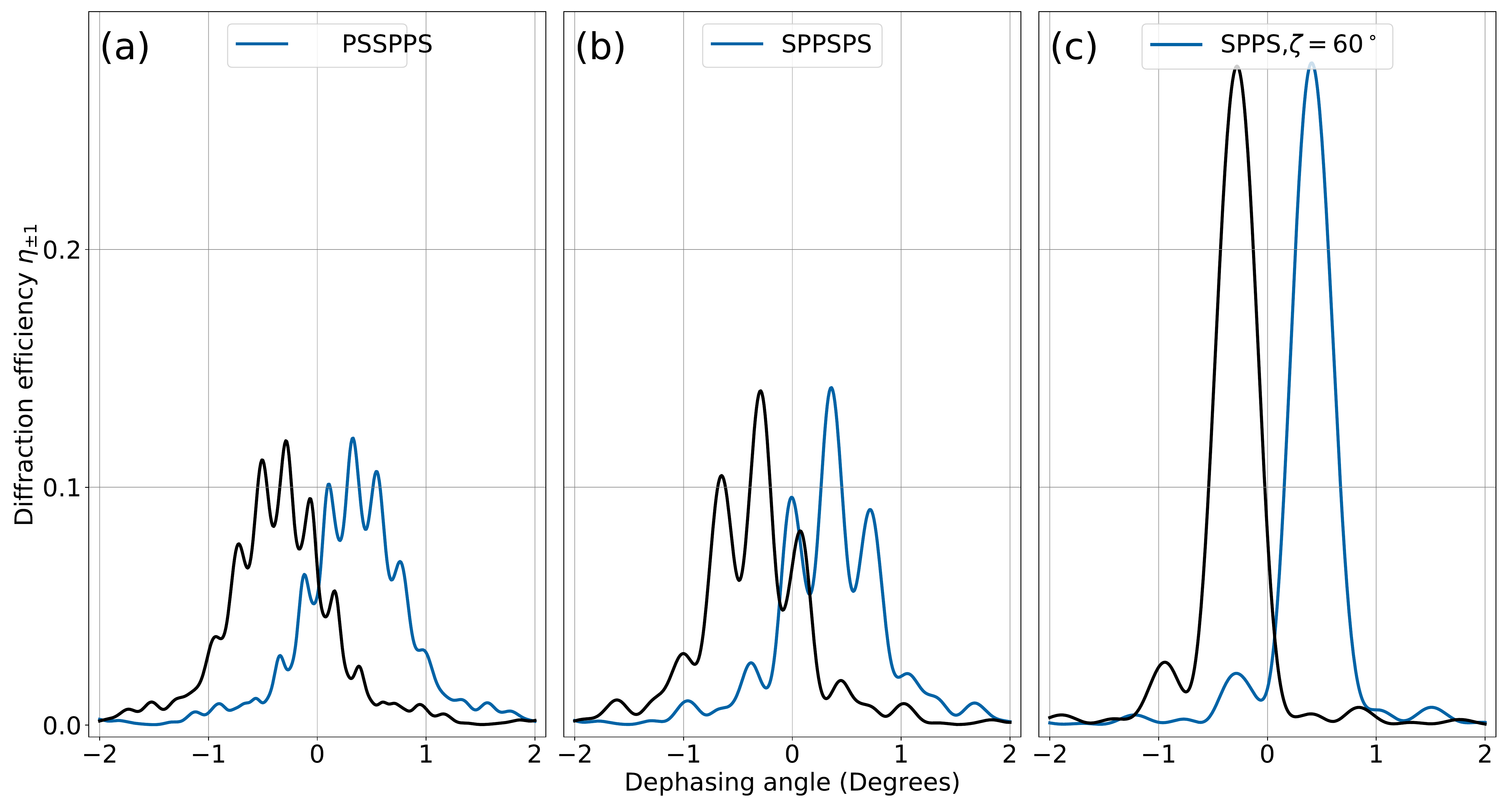}
   \end{tabular}
   \end{center}
   \caption[light1Lred] 
   { \label{fig:sim_3F} 
 (a) Simulated angular dependence of the diffraction efficiency $\eta_{\pm 1}(\Theta)$ for neutrons for a configuration \textsf{PSSPPS} and (b) \textsf{SPPSPS}. The case of \textsf{SPPS} for a tilt angle of $\zeta=60^\circ$ and thus an effective doubling of each layer thickness is shown in (c) for comparison.}
   \end{figure}

For simulations of the other two configurations 3 and 4 we obtain the rocking curves shown in Fig.~\ref{fig:sim_3F}(a),(b).

Finally, gratings can be tilted around the grating vector by about, say, $\zeta=60^\circ$. Then the thickness is doubled since $d\rightarrow d(\zeta)=d/\cos(\zeta)$ \cite{Fally-prl10,Klepp-pra11,Klepp-Materials12,Tomita-prappl20}. The outcome for the simple configuration \textsf{SPPS} (configuration \ref{en:spps}) is depicted in Fig.~\ref{fig:sim_3F}(c) for $\zeta=60^\circ$.

\subsection{Summary and Conclusion}
While the light optical properties and figures of merit are excellent, the neutron optical coherent scattering length density modulation $\bcr$ is only average. However, there is a number of tentative advantages of \bay over other materials:

\begin{itemize}
 \item the ease of preparation without elaborate chemical treatment
 \item the resulting grating quality
 \item the high refractive-index modulation (for light)
 \item the high resolution: even at rather short grating spacings $\Lambda$ a considerable reduction of $n_1$ is not observed. This is particularly important for neutron optic devices.
 \item multilayer stacks can be prepared
\end{itemize}

For the $\textsf{SPPS}$ configuration and light we find a thickness $d_G$ which is slightly less than the nominal values of $2\times 16=32~\mu\textrm{m}$. The grating strength $\nu=n_1\pi d_G/(\lambda_r\cos(\theta_B))\approx 2.8$ is overcoupled ($\nu>i/2$) even for the small thickness delivering a diffraction efficiency $\eta(\Theta_B)=\sin^2\nu\approx 0.2$ far in the second quadrant.

The good news concerning the neutron properties is that \bay gratings do diffract neutrons. However, the coherent neutron scattering length density modulation $\bcr\approx 3~\mu\textrm{m}^{-2}$ for this particular grating is only in the intermediate range as compared to other already explored materials \cite{Fally-prl06,Fally-prl10,Klepp-pra11,Klepp-apl12.01}. We anticipate that tuning and subsequently further improving the recording process will lead to substantially higher values. 

The $\textsf{SPSP}$ configuration for light shows that the angular dependence of the diffraction efficiency oscillates with a frequency dependent on the thickness of the substrate layer, the envelope on the thickness of a single grating layer. According to Eq.~(\ref{eq:2CW-3L1}) values around the Bragg peak might reach diffraction efficiencies of unity. Therefore, we see the following intriguing aspects for using such stacked gratings for neutrons: (1) the rapid oscillations allow to use them as wavelength filters, (2) the wavelength spectra can be estimated, and (3) the angular range of high off-Bragg diffraction efficiencies is extended to the width of the rocking curve of a single layer.

\acknowledgments 
 
We are grateful to Dr. Bruder and Covestro AG for providing sheets of various \bay films. The hospitality of the ILL and technical assistance by Thomas Brenner at the very cold neutron beamline PF2/VCN is acknowledged. 

This research was funded in part by the Austrian Science Fund (FWF) [P-35597-N] and the Austrian Research Promotion Agency (FFG), Quantum-Austria NextPi, grant number FO999896034.

\bibliography{/home/fallym4/texmf/bibtex/bib/misc/isistr,/home/fallym4/texmf/bibtex/bib/misc/publications_url,/home/fallym4/texmf/bibtex/bib/misc/neutron,/home/fallym4/texmf/bibtex/bib/misc/holo,/home/fallym4/texmf/bibtex/bib/misc/bayfol,report} 

\end{document}